\documentclass[12 pt]{article}

\usepackage[dvips]{graphicx}

\def\fo{\hbox{{1}\kern-.25em\hbox{l}}}

\def\beq{\begin{equation}} \def\eeq{\end{equation}}
\def\eq{\end{equation}}

\newcommand{\newc}{\newcommand}
\newc{\gsim}{\lower.7ex\hbox{$\;\stackrel{\textstyle>}{\sim}\;$}}
\def\beq{\begin{equation}} \def\eeq{\end{equation}}

\def\slashchar#1{\setbox0=\hbox{$#1$}           
   \dimen0=\wd0                                 
   \setbox1=\hbox{/} \dimen1=\wd1               
   \ifdim\dimen0>\dimen1                        
      \rlap{\hbox to \dimen0{\hfil/\hfil}}      
      #1                                        
   \else                                        
      \rlap{\hbox to \dimen1{\hfil$#1$\hfil}}   
      /                                         
   \fi}                                         %
\catcode`@=11
\long\def\@caption#1[#2]#3{\par\addcontentsline{\csname
  ext@#1\endcsname}{#1}{\protect\numberline{\csname
  the#1\endcsname}{\ignorespaces #2}}\begingroup
    \small
    \@parboxrestore
    \@makecaption{\csname fnum@#1\endcsname}{\ignorespaces #3}\par
  \endgroup}

\usepackage{feynmf}
\begin{document}

\begin{titlepage}

\begin{flushright}
UCD-2000-16   \\ 
LBNL-47049
\end{flushright}

\huge
\begin{center}
{\Large\bf Nucleon scattering with higgsino \\
and wino cold dark matter}
\end{center}

\large

\vspace{.15in}
\begin{center}

Brandon Murakami${}^a$ and James D.~Wells${}^{a,b}$

\small

\vspace{.1in} {\it ${}^{(a)}$Davis Institute of High Energy Physics \\
University of California, Davis, California 95616 \\}
{\it ${}^{(b)}$Lawrence Berkeley National Laboratory, Berkeley, CA 94720 }

\end{center}
 
\vspace{0.15in}
 
\begin{abstract}

Neutralinos that are mostly wino or higgsino are shown to be
compatible with the recent DAMA annual modulation signal.  The
nucleon scattering rates
for these dark matter candidates are typically an order of magnitude
above the oft-considered bino.  Although thermal evolution of higgsino
and wino number densities in the early universe implies that they are
not viable dark matter candidates, non-thermal sources, such as from
gravitino or moduli decay in anomaly mediated supersymmetry breaking, 
suggest that they can be the dominant
source of cold dark matter.  Their stealthiness at high energy colliders
gives even more impetus to analyze nucleon scattering detection
methods.  We also present calculations for their predicted scattering rate
with germanium detectors, which have yet to see evidence of WIMP scattering.

\end{abstract}

\begin{flushleft}
hep-ph/0011082 \\
November 2000
\end{flushleft}

\end{titlepage}

\baselineskip=18pt


\newcommand{\lsp}{{\tilde\chi}} \newcommand{\squark}{{\tilde q}}
\newcommand{\lag}{{\begin{mathcal} L \end{mathcal}}}
\section{Introduction}
\bigskip

There are two intriguing reasons to study the lightest neutralino from
supersymmetric extensions of the Standard Model.  For one, we would
like an answer for what composes the non-baryonic matter necessary to
explain the observed rotations of galaxies.  By angular
momentum considerations, we note that the observed visible matter in
our galaxy implies a lower theoretical angular velocity than observed.
We also observe galaxies in clusters insufficiently bound by 
the visible matter. Secondly, we demonstrate in this
article that should light neutralinos exist as a cold dark matter
(CDM) candidate, the detection prospects are promising for a wide
range of parameter space, including for higgsinos and winos.

In the various supersymmetric models, the sneutrino, the gravitino,
and a neutralino are all capable of being the lightest supersymmetric
particle (LSP) \cite{ellis84}.  Here, we study the lightest neutralino
$\lsp$ as the LSP.  Depending on the supersymmetry breaking model
considered, the lightest neutralino may be bino-like, wino-like,
higgsino-like, or admixtures of these three categories.

Historically, the bino-like LSP has been
studied extensively as a dark matter candidate~\cite{diehl95}.
The wino-like and higgsino-like LSP have been largely ignored as dark
matter candidates
because they can be demonstrated in the majority of popular models, that
their thermal relic densities are simply too low for dark matter contention
\cite{mizuta93}.

We include the possibility of a relatively new class of models --- the
anomaly-mediated supersymmetry breaking (AMSB) scenarios
\cite{randall99, giudice98, bagger00}.  The relevant characteristic of
these models is the possibility of a wino-like or higgsino-like LSP
(from a non-thermal
source) with a density capable of being the dark matter
\cite{gherghetta99, moroi00}.  The wino-like LSP is not unique to the
AMSB scenarios.  Theoretical grounds are also provided by
moduli-dominated supersymmetry breaking within O-II superstring
motivated models \cite{brignole95} and supersymmetry breaking in which
the responsible $F$-term is not a SU(5) singlet \cite{snowmass96}.
What is unique about AMSB is its prediction that the gravitino mass
is a few orders of magnitude above the superpartner masses.  This
characteristic is what ultimately enables the wino or higgsino to
be a viable dark matter candidate.

If the lightest neutralino truly is the dark matter or a significant
fraction, they, being weakly interacting in character, will penetrate
the earth much like a neutrino.  This allows for possible detection by
scattering elastically with nucleons.  Of the different neutralino LSP
detection methods~\cite{KamGriest}, we report on the detection prospects
of the lightest neutralino scattering off nucleons within different target
samples.

\section{Supersymmetry Breaking Models}
\bigskip

\newcommand{\tchi}{\tilde\chi}

First, a brief review of how the neutralino obtains its character is
supplied.  The Standard Model includes an electrically neutral U(1)
gauge field $B$ and a neutral SU(2) gauge field $W^3$.  The minimal
supersymmetric extension includes two neutral Higgs scalar fields,
$H_d^0$ and $H_u^0$.  These four fields have supersymmetric fermionic
partners $\tilde\psi = (\tilde{B}, \tilde{W}^3, \tilde{H}_d^0,
\tilde{H}_u^0)$ with various interactions amongst one another,
allowing for a non-diagonal mass matrix to be formed out of the
Lagrangian quadratic terms.  A transformation may be performed to
diagonalize the neutralino mass matrix to the physical basis, $\tchi =
(\tchi_1, \tchi_2, \tchi_3, \tchi_4)$, with $\tchi_1$ having the
smallest mass eigenvalue.  The Haber and Kane convention~\cite{HaberKane} 
for $\mu$
is used in the mass
matrix.  The notation $\lsp$ will be used to mean the lightest
neutralino ($\tchi_1$).  If, for example, the largest component of
$\lsp$ is the $\tilde W^3$ component, we say the LSP is wino-like.
Thus, the LSP's eigenvector components determine its character, as
made explicit by

\beq \lsp = N_{11} \tilde B + N_{12} \tilde W^3 + N_{13} \tilde H_d +
N_{14} \tilde H_u.  \eeq

We consider gravity mediated and anomaly mediated supersymmetry
breaking.  The values of $M_1$ and $M_2$ are determined within these
models, with $\mu$ constrained only by additional assumptions on the
superpartner spectrum.  In a large class of minimal
supergravity supersymmetry breaking scenarios, scalar masses $m_0$ 
and gaugino masses $M_i$ are
chosen to be the same at the unification scale.  In these models, the
relation between the U(1) and SU(2) gaugino masses is

\begin{eqnarray} 
\frac{M_1}{g_1^2} = \frac{M_2}{g_2^2} & (m_{1/2} \textrm{ universal at
GUT scale}).
\end{eqnarray}

\noindent
This amounts to $M_1 = \frac{5}{3} \tan^2\theta_W M_2$ being about
half $M_2$ at the electroweak scale.  Within these models, there is
still a chance that $\mu$ is smaller than $M_1$.  The
lightest neutralino $\lsp$ will then be either bino- or higgsino-like
for these models.

A wino-like $\lsp$ is possible within the AMSB class of models.  AMSB
models supply two results relevant to our discussion.  The first of
which claims that the gaugino masses are proportional to the gauge
coupling beta function,
\begin{eqnarray}
M_n = \frac{\beta_{g_n}}{g_n} m_{3/2} & \textrm{(AMSB)}
\end{eqnarray}
where $g_n$ is the gauge coupling constant, $m_{3/2}$ is the gravitino
mass, and $\beta_{g_n}$ is the beta function of the gauge coupling
constant.  At the electroweak scale, the values of $g_n$ and
$\beta_{g_n}$ result in $M_2$ being approximately a third of $M_1$.
Furthermore, $\mu < M_2$ is possible, depending on 
the sparticle spectrum.
In this case, AMSB would produce a higgsino LSP.

The wino's cross-section for pair annihilation is higher than the
bino's for two reasons:  $g_2$ is larger than $g_1$, and bino pair
annihilation has less options for the final state than winos.  For example,
winos can annihilate to $W$ pairs without any intermediate scalar 
superpartners whereas pure binos cannot.
This results in a present day number
density $n_\lsp$ too low for winos as a cold dark matter candidate
(but often just right for binos, which originates its appeal as a cold
dark matter candidate).  This rules out a wino-like LSP and is the
reason the wino-like LSP has been largely ignored as a CDM candidate.
Let us call this a ``thermal source'' for wino production.  This
disucssion also applies to an AMSB-produced higgsino-like LSP.

AMSB supplies a second, ``non-thermal source'' for $\lsp$ production.
The gravitino or moduli fields in this model are
what constitute the second LSP source via direct or indirect
$\lsp$ decay products.
A modulus field may be defined as the scalar field that parameterizes
a flat direction in the theory.  The
moduli fields may acquire a mass from supersymmetry breaking as the
moduli's own potentials are ``lifted'' at the ends.
The important feature of AMSB models is the granting of a large
mass (10 to 100 TeV) to the gravitino and moduli fields.  These large
masses can circumvent the historically annoying gravitino
problem and  ``cosmological moduli problem'' in which the gravitino
and  moduli acquire small masses under common supergravity models.  A
small enough mass may have a lifetime longer than successful Big Bang
nucleosynthesis would allow.  In general, it is undesirable for
any cosmological scenario to include a field dumping Plank scale
energy into the universe after nucleosynthesis begins.  The large
masses from AMSB models ensure the gravitino and moduli decay
sufficiently early \cite{gherghetta99, moroi00}.

\section{The Scattering Rate}
\bigskip

\begin{figure}[!btp]
\begin{center}
\begin{tabular}{cccc}
\begin{fmffile}{hexchange}
\begin{fmfgraph*}(65,60)
\fmftop{chiin,chiout} \fmfbottom{qin,qout}
\fmf{fermion}{chiin,v1,chiout} \fmf{dashes,label=$h$}{v1,v2}
\fmf{fermion}{qin,v2,qout} \fmflabel{$\lsp$}{chiin}
\fmflabel{$\lsp$}{chiout} \fmflabel{$q$}{qin} \fmflabel{$q$}{qout}
\fmfdot{v1,v2}
\end{fmfgraph*}
\end{fmffile}
\quad\quad &
\begin{fmffile}{Zexchange}
\begin{fmfgraph*}(65,60)
\fmftop{chiin,chiout} \fmfbottom{qin,qout}
\fmf{fermion}{chiin,v2,chiout} \fmf{boson,label=$Z$}{v1,v2}
\fmf{fermion}{qin,v1,qout} \fmflabel{$\lsp$}{chiin}
\fmflabel{$\lsp$}{chiout} \fmflabel{$q$}{qin} \fmflabel{$q$}{qout}
\fmfdot{v1,v2}
\end{fmfgraph*}
\end{fmffile}
\quad\quad &
\begin{fmffile}{tsqexchange}
\begin{fmfgraph*}(65,60)
\fmftop{chiin,qout} \fmfbottom{qin,chiout}
\fmf{fermion}{chiin,v1,qout} \fmf{dashes,label=$\squark$}{v1,v2}
\fmf{fermion}{qin,v2,chiout} \fmflabel{$\lsp$}{chiin}
\fmflabel{$\lsp$}{chiout} \fmflabel{$q$}{qin} \fmflabel{$q$}{qout}
\fmfdot{v1,v2}
\end{fmfgraph*}
\end{fmffile}
\quad\quad &
\begin{fmffile}{qloops}
\begin{fmfgraph*}(65,60)
\fmftop{chiin,chiout} \fmfbottom{gin,gout}
\fmf{fermion}{chiin,v,chiout} \fmf{gluon}{gin,v} \fmf{gluon}{v,gout}
\fmfblob{0.2w}{v} \fmflabel{$\lsp$}{chiin} \fmflabel{$\lsp$}{chiout}
\fmflabel{$g$}{gin} \fmflabel{$g$}{gout}
\end{fmfgraph*}
\end{fmffile}
\\[5mm] \small{(a) $h$- and $H$-exchange} &\small{(b) $Z$-exchange}
&\small{(c) $\squark$-exchange} &\small{(d) scattering with gluons}\\
&&& \small{via heavy quark loops}
\end{tabular}
\caption{The leading order parton diagrams for $N-\lsp$ scattering.}
\end{center}
\label{diagrams}
\end{figure}
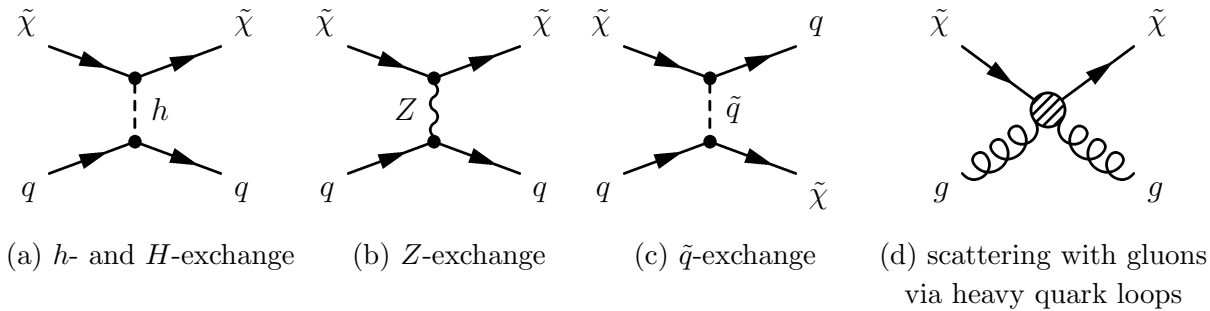

The detection prospects of a neutralino from the galactic halo are
quantified as the elastic scattering (or event) rate $R$.  The
assumed virialized neutralinos are bound to the 
galactic halo by a gravitational
potential, and the scattering rate (measured in
events/day/kg) is \cite{drees93oct}

\beq R = \frac{\sigma \rho_\lsp v_\lsp F_\xi}{m_\lsp M_N}.  \eeq

\noindent
where $\rho_\lsp = m_\lsp n_\lsp$ is the mass density of the
neutralinos, $v_\lsp$ is the average speed of the neutralinos as they
float around the galactic halo, $M_N$ is the mass of the target
nucleus, and $F_\xi$ is the nuclear form factor
(we take $F_\xi=1$).  In terms
of the assumed average values, the neutralino's speed and density, the
scattering rate is
\beq 
R = \frac{\sigma F_\xi}{m_\lsp M_N} 1.8 \times 10^{11}
\mathrm{GeV}^4 \left( \frac{\rho_\lsp}{0.3 \mathrm{GeV/cm}^3} \right)
\left( \frac{v_\lsp}{320 \mathrm{km/sec}} \right) \left(
\frac{\textrm{events}}{\mathrm{day} \cdot \mathrm{kg}} \right).  
\eeq

The remainder of this section is organized as follows:  our model's
lagrangian is stated, the possible processes are discussed, and the
many coupling constants are made explicit.
The Lagrangian is divided into two parts:  spin-dependent
terms and spin-independent terms.  This separation is motivated by
realizing  spin-independent interactions are enhanced by the presence
of all the nucleons in the nucleus.

To describe the spin-independent interactions, we use an effective
Lagrangian interaction term of the form
\beq 
\lag \subset \sum_{N\in\{p,n\}} f_N[\bar{\lsp}\lsp][\bar{N}N],
\eeq
where the subscript $N$ is the nucleon field, proton or neutron.  $f_N$ is
the scalar 4-point effective coupling constant that includes Higgs and
squark exchange, as well as the neutralino-gluon scattering of 
Fig.\ 1(d).  The contribution of scattering with gluons occurs through a
loop of quarks or squarks and is fully calculated in
Ref.\ \cite{drees93may}.   It is shown to be at least an order of
magnitude smaller than Higgs exchange in Ref.\ \cite{drees93oct} for
most of parameter space involving reasonably heavy squarks.  For this
reason, only one neutralino-gluon interaction is large enough for our
task --- Higgs exchange to a triangle of heavy quarks ($c$, $b$, $t$)
coupled to gluons.  These interactions are in addition to another
effective Lagrangian term that describes spin-dependent $Z$, Higgs,
and squark exchange:
\beq 
\lag \subset [\bar\lsp \gamma^\mu \gamma^5 \lsp] [\bar q
\gamma_\mu(c_q + d_q \gamma^5)q] 
\eeq
where $c_q$ and $d_q$ are effective coupling constants to be made
explicit shortly.

These Lagrangian terms yield a scattering cross-section of
\cite{drees93oct}
\beq 
\sigma = \frac{4}{\pi} \left(\frac{m_\lsp M_N}{m_\lsp + M_N}
\right)^2 \left[ (n_p f_p + n_n f_n)^2 + 4 \lambda^2 J(J+1) \left(
\sum_{q \in \{u,d,s\}} d_q \Delta q \right)^2 \right].
\eeq
The factor $\lambda^2 J(J+1)$ represents the fraction of the nucleons'
spin in comparison to the total nucleus spin (squared) $J(J+1)$, which
includes orbital angular momentum.  $m_\lsp$, $n_p$, and $n_n$ are the
masses of the lightest neutralino, a proton, and a neutron,
respectively.  $\Delta q$ is a quark's second moment of the quark
density of its spin polarization \cite{ellis93}.

The spin-independent effective coupling constant $f_N$ is composed of
two terms:

\beq f_N = f_H + f_D. \eeq

\noindent
The spin-independent squark exchange contributions are organized into
terms proportional to the sums and differences of MSSM Yukawa coupling
constants $a_{\squark_i}$ and $b_{\squark_i}$ from the $\squark_i q
\lsp$ vertex, supplied in the Appendix.  The terms proportional to
$a_{\squark_i}-b_{\squark_i}$ are represented by $f_D$. Those
proportional to $a_{\squark_i}+b_{\squark_i}$ are small and omitted
\cite{drees93oct}.  $f_H$ is the spin-independent contribution due to
Higgs exchange to all six flavors of quarks.  The form of $f_H$ is

\beq f_H = \sum_{q \in \{u,d,s\} } \frac{f_q^{H}}{m_q} f_{Tq}m_N +
\frac{2}{27} \sum_{q \in \{c,b,t\} } \frac{f_q^{H}}{m_q} f_{Tg}m_N \eeq

\noindent
where $f_{Tq}$ is the fraction of the nucleon mass the light quarks
($u$, $d$, $s$) effectively represent, defined by $f_{Tq}m_N \equiv
\langle N|m_q\bar{q}q|N\rangle$ for nucleon $N$.  \mbox{$f_{Tg} \equiv 1
- \sum f_{Tq}$} is the remaining fraction carried by the nucleon sea.
$f_q^H$ is the effective coupling constant for a $f_q^H [\bar{\lsp}
\lsp][\bar{q}q]$ Lagrangian term that describes Higgs exchange.
$f_q^H$ follows from a straightforward MSSM calculation of $h$ and
$H$ $t$-channel exchange between $\lsp$ and a quark:

\beq f_q^H = m_q \Bigg( \frac{c_{h \tilde\chi \tilde\chi} c_{hqq}
}{m_h^2} + \frac{c_{H \tilde\chi \tilde\chi} c_{Hqq} }{m_H^2} \Bigg)
\label{fqH}
\eeq

\noindent
where the MSSM Yukawa coupling constants for Higgs-neutralino and
Higgs-quark are supplied in the Appendix.

The squark exchange interactions proportional to $a_{\squark_i}^2 -
b_{\squark_i}^2$ have an effective coupling of

\beq f_D = \sum_{q \in \{u,d,s\} } \frac{f_q^\squark}{m_q} f_{Tq} m_N.
\eeq

\noindent
This refers to simple $s$-, $t$-, and $u$-channel exchanges with the
light quarks.  All squark exchange interactions proportional to
$a_{\squark_i}^2 - b_{\squark_i}^2$, denoted $f_S$, are at least an
order of magnitude smaller than the total interaction $f$, even for
reasonably light squarks.

Squark exchange appears as both spin-dependent and spin-independent.
The 4-point Lagrangian interaction term
implied by $s$-, $t$-, and $u$-channel diagrams of neutralino-quark
scattering by squark exchange is $[\bar{q}\lsp][\bar{\lsp}q]$.  The
Fierz rearrangement supplies the scalar form

\nonumber \beq \lag \subset f_q^\squark[\bar{\lsp}\lsp][\bar{q}q] \eeq

\noindent
as well as the vector/axial-vector form

\nonumber \beq \label{eq:spinindL} \lag \subset
[\bar{\lsp}\gamma^\mu\gamma^5\lsp][\bar{q}\gamma_\mu(c_q^\squark +
d_q^\squark \gamma^5)q], \eeq

\noindent
where $f_q^\squark$, $c_q^\squark$, and $d_q^\squark$ are effective
coupling constants:

\begin{eqnarray}
f_q^\squark &=& - \frac{1}{4} \sum_{i=1}^2 \frac{a_{\squark_i}^2 -
b_{\squark_i}^2}{m_{\squark_i}^2 - (m_\lsp + m_q)^2 } \\ c_q^\squark
&=& - \frac{1}{2} \sum_{i=1}^2 \frac{a_{\squark_i}
b_{\squark_i}}{m_{\squark_i}^2 - (m_\lsp + m_q)^2 } \\ d_q^\squark &=&
\frac{1}{4} \sum_{i=1}^2 \frac{a_{\squark_i}^2 +
b_{\squark_i}^2}{m_{\squark_i}^2 - (m_\lsp + m_q)^2 }.
\end{eqnarray}

\noindent
$a_{\squark_i}$ and $b_{\squark_i}$ refer to the MSSM couplings for
the $\squark q\lsp$ vertices and are supplied in the Appendix.

As for spin-dependent interactions, $Z$-exchange and squark-exchange
involve a vector coupling constant $c_q$ and an axial vector coupling
constant $d_q$:
\begin{eqnarray}
c_q = c_q^\squark+c_q^Z, & & d_q = d_q^\squark+d_q^Z.
\end{eqnarray}
$c_q^Z$ and $d_q^Z$ are effective coupling constants for a 4-point
$\lsp\lsp qq$ vertex that implicitly proceeds through $Z$-exchange:
\begin{eqnarray}
c_q^Z = \frac{g_2^2 O''_R (T_{q3} - 2 e_q \sin^2\theta_W)} {4m_W^2}, &
& d_q^Z = -\frac{g_2^2 O''_R T_{q3}}{4 m_W^2} \label{dZq}
\end{eqnarray}
where $T_{q3}$ is the quark's weak isospin third component, $e_q$ is
the electric charge of the quark, and $O''_R$ is the $Z\lsp\lsp$
vertex coupling from the MSSM (see the Appendix).

\begin{figure}[!btp]
\begin{tabular}{ccc}
\includegraphics{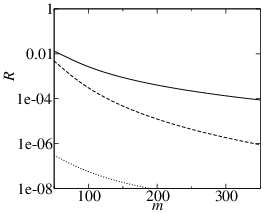}& \includegraphics{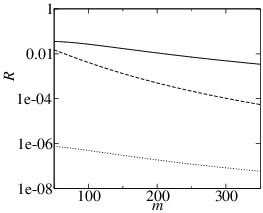}&
\includegraphics{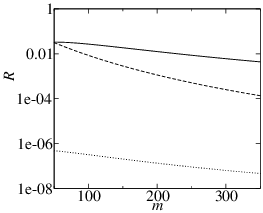}\\ \small{(a) $\tilde{B}$-like $\lsp$}
&\small{(b) $\tilde{W}$-like $\lsp$} &\small{(c) $\tilde{H}$-like
$\lsp$}
\end{tabular}
\caption{Contributions to $R$ (events/day/kg) vs.\ $m_\chi$ (GeV).
These are the contributions to the $\lsp$-$^{73}$Ge scattering rate $R$
by Higgs (solid), squark (dotted), and $Z$ exchange (dashed).  These
contributions are calculated from (a) a mSUGRA bino-scenario
($M_1=\frac{1}{3}\mu$, $M_2=\frac{2}{3}\mu$) , (b) an AMSB
wino-scenario ($M_1=\frac{3}{2}\mu$, $M_2=\frac{1}{2}\mu$), and (c) a
higgsino-scenario ($M_1=\frac{3}{2}\mu$, $M_2=3\mu$).  The
remaining parameters are $\tan\beta=4$, squark masses $m_\squark=2$
TeV, and soft trilinear parameters $A_\squark=0$, pseudo-scalar Higgs
mass $m_A=500$ GeV, and the nuclear and astrophysical parameters of
Table 1.}
\end{figure}

\section{Scattering off Germanium}

We now numerically compare the contributions to the scattering rate by
the Higgs, squark, and $Z$ exchange in Figures 2.  The target sample
is germanium to match the CDMS \cite{Abusaidi:2000wg} and HDMS
\cite{Baudis:2000ph} collaborations.
The nuclear and astrophysical parameters used are
listed in Table 1.  The nuclear parameters are from the 1988 EMC data
\cite{EMC89,cheng89,Jaffe90} 
and Ref.\ \cite{ellis91}.  For reasonable parameter
constraints supplied by AMSB and mSUGRA, we created scenarios with the
lightest neutralino as wino-like, bino-like, and higgsino-like.  From
the numerical results, the Higgs-exchange contribution is seen to be
much larger than squark-exchange and $Z$-exchange.  The contribution from $Z$-exchange becomes significant for low values of $\mu$ (which is roughly proportional to $m_\lsp$) due to the cross-section contribution of the form $m_r/\mu^4$ where $m_r$ is the reduced mass of the nucleus and neutralino.

To analytically demonstrate the Higgs dominance, first consider the
effective coupling constant $f_q^H$ of the 4-point interaction that
proceeds through Higgs exchange.  It is proportional to the MSSM
$h\lsp\lsp$ and $H\lsp\lsp$ vertices.  In the approximations of
wino-like lightest neutralino ($N_{12} \gg
N_{11}$) and small Higgs mixing angle $\alpha$, these
coupling constants are

\begin{table}
\caption{\small{Nuclear and astrophysical parameters used for
calculations \cite{EMC89,cheng89,Jaffe90, ellis91}.}}
\begin{center}
\begin{small}
\begin{tabular}{lr}
\hline for protons: \\ $f_{Tu}$ & 0.023 \\ $f_{Td}$ & 0.034 \\
$f_{Ts}$ & 0.14 \\ \hline for neutrons: \\ $f_{Tu}$ & 0.019 \\
$f_{Td}$ & 0.041 \\ $f_{Ts}$ & 0.14 \\ \hline
\end{tabular}
\end{small}
\hspace{0.25 in}
\begin{small}
\begin{tabular}{lr}
\hline $\Delta u$ & 0.77 \\ $\Delta d$ & -0.49 \\ $\Delta s$ & -0.15
\\ \hline for $^{73}$Ge,
$\lambda^2 J(J+1)$ & 0.065 \\ \hline 
$F_\xi$ & 1 \\ 
$v_\lsp$ & 320 km/s \\ 
$\rho_\lsp$ & 0.3 GeV/cm$^3$ \\ \hline
\end{tabular}
\end{small}
\end{center}
\end{table}

\begin{eqnarray}
c_{h\chi\chi} & \approx & \frac{1}{2} g_2 N_{12} N_{14} \\
c_{H\chi\chi} & \approx & -\frac{1}{2} g_2 N_{12} N_{13}.
\end{eqnarray}

\noindent
Further imposing the limits $m_Z/|\mu\pm M_1| \ll 1$, $m_Z/|\mu\pm
M_2| \ll 1$, and $\tan\beta>2$, these coupling constants become \cite{moroi00}

\begin{eqnarray}
c_{h\chi\chi} & \approx &
\frac{g_2}{2} \frac{M_2 + \mu\sin{2}\beta}{\mu^2 - M_2^2} m_W \label{hWW} \\
c_{H\chi\chi} & \approx &
\frac{g_2}{2} \frac{\mu\cos{2}\beta}{\mu^2-M_2^2} m_W. \label{HWW}
\end{eqnarray}

\noindent
For sketching purposes, these couplings go as ${\sim}m_W/\mu$. The
contribution of Higgs-exchange to the cross section will then go as
${\sim}m_r^2 m_N^2/\mu^2 m_h^4$.

The coupling, $d_q^Z$, for $\lsp q$ scattering through $Z$-exchange
(Eqn. \ref{dZq}) goes as ${\sim}m_Z^2/\mu^4$ and contribution to the
cross-section goes as ${\sim}m_r^2 m_Z^4/\mu^8$.  The wino- and
higgsino-like neutralino scenarios of figures 2(b) and 2(c) show over
an order of magnitude $Z$-exchange increase over the bino-like
scenario since their cross-sections are proportional to $g_2^4$,
versus $g_1^4$.

The squark-exchange couplings $f_q^\squark$, $c_q^\squark$, and
$d_q^\squark$ are proportional to $a_{\squark_i}^2 + b_{\squark_i}^2$,
$a_{\squark_i} b_{\squark_i}$, and $a_{\squark_i}^2 -
b_{\squark_i}^2$.  $a_{\squark_i}$ and $b_{\squark_i}$ will be of the
same order for all types of neutralinos if $\tan\beta$ is not
extremely large or small.  An extreme value of $\tan\beta$ may cause
squark-exchange to become more pronounced (see the Appendix for the
$Z_{q0}$ coupling).  For moderate values of $\tan\beta$, it is because
the squark cross-section contribution is proportional
to $\sim m_r^2/m_\squark^4$ that squark exchange is suppresed when a heavy
squark is considered.

\begin{figure}
\begin{tabular}{ccc}
\includegraphics{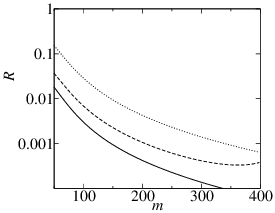}& \includegraphics{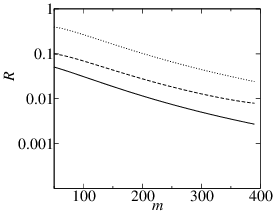}&
\includegraphics{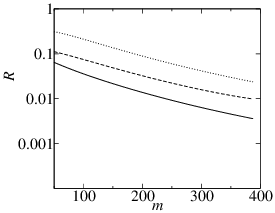}\\ \small{(a) $\tilde{B}$-like $\lsp$}
&\small{(b) $\tilde{W}$-like $\lsp$} &\small{(c) $\tilde{H}$-like
$\lsp$}
\end{tabular}
\caption{The $\lsp$-$^{73}$Ge scattering rates $R$ (events/day/kg) as a
function of the LSP mass $m_\lsp$ (GeV) for bino, wino, and higgsino scenarios.
Scenario (a) uses $M_1=\frac{1}{3}\mu$, $M_2=\frac{2}{3}\mu$, (b) uses
$M_1=\frac{3}{2}\mu$, $M_2=\frac{1}{2}\mu$, and (c) uses
$M_1=\frac{3}{2}\mu$, $M_2=3\mu$.  The solid lines have ``moderate''
parameters $m_\squark=2$ TeV, $m_A=500$ GeV.  The dashed lines
represent a light squark with $m_\squark=500$ GeV, $m_A=500$ GeV.  The
dotted lines represent a light pseudo-scalar Higgs with $m_\squark=2$
TeV, $m_A=150$ GeV.  For all, $\tan\beta=4$ and soft trilinear
parameters $A_\squark=0$ are used.}
\end{figure}

Fig.\ 3 shows the numerically calculated scattering rates for
neutralinos and $^{73}$Ge at $\tan\beta=4$ for the three cases of
``moderate'' values of the squark mass ($m_\squark=2$ TeV) and
pseudo-scalar Higgs masses ($m_A=500$ GeV), a light squark
($m_\squark=500$ GeV) case, and a light pseudo-scalar Higgs ($m_A=150$
GeV) case.  The practical aspect of these plots lies in recognizing
the various possibilities of tuning the cross-section and scattering
rate by varying the supersymmetry breaking model parameters $\mu$,
$m_{\squark_i}$, and $m_A$ along with $\tan\beta$.  The
rates for $^{76}$Ge are similar to those for $^{73}$Ge since the
spin-dependent contributions are small for both.

The bino-like neutralino scattering rate suffers from the lower $g_1$
coupling constant, explicit in Eqs.\ \ref{hXX} and \ref{HXX}.  The
wino- and higgsino-like neutralinos have comparable scattering rates
due to the common vertices that originate from the Lagrangian terms
$-\frac{1}{\sqrt{2}} g_2 H_i \tilde{H}_i \tilde{W}^3$.
The bino-like neutralino has analogous terms but are proportional to the
smaller U(1) gauge coupling $g_1$.  Analogous to Eqs. \ref{hWW} and
\ref{HWW}, using the same limits, the higgsino-like LSP from mSUGRA has the
Yukawa couplings

\begin{eqnarray}
c_{h\lsp\lsp} &\approx& -\frac{g_2}{4} \frac{1+\sin 2\beta}{\mu-M_1} m_W \tan\theta_W, \\
c_{H\lsp\lsp} &\approx& -\frac{g_2}{4} \frac{2\sin^2\beta-1}{\mu-M_1} m_W \tan\theta_W.
\end{eqnarray}

While these calculations show promising event rates for a wide range
of parameter space, the regions of parameter space that result in
undetectable scattering rates should be noted.  Larger values of
$\tan\beta$ will lower the $h\bar tt$ coupling (Eq.\ \ref{huu}) and
suppress the light Higgs exchange.  Large $\tan\beta$ also slightly
increases the light Higgs mass $m_h$ and causes further suppression.
In the case of light squarks, a large $\tan\beta$ has the opposite
effect --- enhancing the scattering rate by causing significant
left-right squark mixing, driving the sbottom mass down.
The corresponding squark propagators are then enhanced.  Overriding
the effects of varying $\tan\beta$, larger mass parameters for the
Higgs, gauginos, and squarks can quickly reduce the scattering rate to
undetectable levels.
When $\tan\beta\gg 1$ the cross-section begins to rise again. One can
show that the cross-section starts to rise like $\tan^2 \beta$
when $\tan\beta \gsim m_H^2/m_h^2$ due to the increased relative importance
of the heavier Higgs boson.

\section{DAMA Constraints on Supersymmetry Parameters}

\begin{figure}[!tbp]
\begin{tabular}{ccc}
\includegraphics{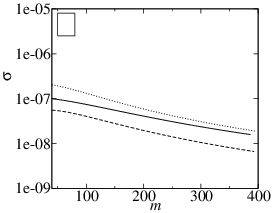}& \includegraphics{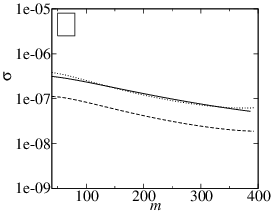}&
\includegraphics{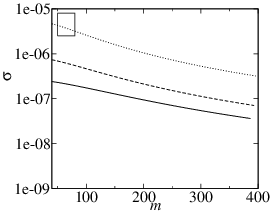}\\ \small{(a) ``moderate'' $m_\squark$
and $m_A$ ($\mu>0$)} &\small{(b) light squark} ($\mu>0$) &\small{(c)
light pseudo-scalar Higgs ($\mu>0$)}\\[2 mm]
\includegraphics{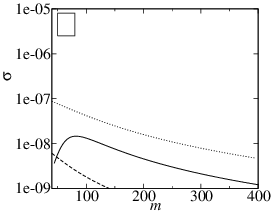}& \includegraphics{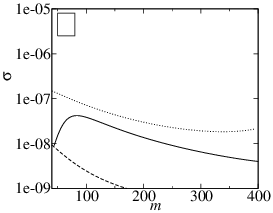}&
\includegraphics{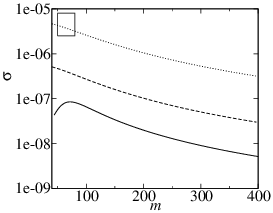}\\ \small{(a') ``moderate'' $m_\squark$
and $m_A$ ($\mu<0$)} &\small{(b') light squark ($\mu<0$)} &\small{(c')
light pseudo-scalar Higgs ($\mu<0$)}
\end{tabular}
\caption{Illustrations of the 
DAMA constraints (roughly equivalent to the boxes in each figure panel).  
Each plot represents the scalar
(spin-independent) contribution to the cross-section $\sigma_{\rm
scalar}$ (pb) as a
function of the wino-like neutralino mass $m_\lsp$ (GeV) for neutralino-proton
scattering in an AMSB scenario ($M_1=\frac{3}{2}\mu$,
$M_2=\frac{1}{2}\mu$).  (a) represents ``moderate'' parameters
($m_\squark=2$ TeV, $m_A=500$ GeV), (b) represents a light squark
scenario ($m_\squark=500$ GeV, $m_A=500$ GeV), and (c) represents a
light pseudo-scalar Higgs scenario ($m_\squark=2$ TeV, $m_A=200$ GeV).
$A_\squark=0$ is used for all soft trilinear parameters.  The values
of $\tan\beta$ are 2 (solid), 10 (dashed) and 30 (dotted).
The upper (lower) set of plots have values $\mu>0$ ($\mu<0$).}
\end{figure}

The DAMA collaboration reports~\cite{bernabei00} 
indications of a WIMP with a mass of
$52^{+10}_{-8}$ GeV and a cross-section of
$\xi\sigma_{\textrm{scalar}} = 7.2^{+0.4}_{-0.9} \times 10^{-6}$ pb
when scattered against a single proton, using standard astrophysical
assumptions (as in Table 1).  $\xi$ is the fraction
of the cold dark matter that this WIMP represents.  The ``scalar''
subscript on the cross-section refers to only the spin-independent
contributions.

We apply our analysis to interpret this WIMP signal as a scattering of
the lightest neutralino.  While the CDMS collaboration claims to exclude the
neutralino parameter space implied by DAMA, the DAMA limits are used here at
face value \cite{Abusaidi:2000wg}.  The AMSB wino scenario is chosen as the
framework for this interpretation without too much loss of generality.
One can roughly equate this with the higgsino scenario, or take
off an order of magnitude from the scattering rate to imagine the
mSUGRA bino scenario (as in Fig.\ 3).

We have made assumptions about the astrophysical and squark parameters
($A_\squark$ and $m_\squark$).  Altering the astrophysical models,
i.e.\ the galactic halo structure, may significantly change the allowed
SUSY parameter space.  Refs.\ \cite{kamionkowski98,
brhlik99,green00} demonstrate a means for different galactic halo models
to allow for WIMP masses up to 150 GeV at $3\sigma$.  Non-universal
squark masses and soft trilinear parameters allow for a broad range
of cross-sections for any fixed set of values $m_\lsp$, $\tan\beta$,
and $\mathrm{sgn}(\mu)$, as seen in Refs.\ \cite{berezinsky95, ellis00,
corsetti00, accomando00}.
As for assumptions on the nuclear parameters, Ref.\ \cite{bottino00}
demonstrates the effect of uncertainties of the quark masses and mass
fractions $f_{Tq}$.

Using the ``moderate'' SUSY parameters of Fig.\ 4(a) ($m_\squark=2$
TeV, $m_A=500$ GeV, $A_\squark=0$) the cross section is significantly
below the favored region of the DAMA/NaI-2 data.  
The lighter squark of Fig.\ 4(b) will increase the scalar cross-section,
but not to DAMA heights.  A light pseudo-scalar Higgs is necessary
for compatibility with the DAMA limits.  By tuning
$\tan\beta$, $m_\squark$, and $m_A$ the DAMA interpretation may be
made compatible with anomaly and gravity mediated SUSY breaking models.

One concern with the above parameter space with light pseudo-scalar is
that the lightest CP-even Higgs mass will fall below the LEP-II limits.
Indeed, our lightest Higgs mass in our examples range from about
100 GeV to 115 GeV which is just at the LEP-II SM Higgs limits.
However, one should keep in mind that the LEP-II SUSY Higgs limits are
significantly below the SM Higgs limits due to decreased $hZZ$ coupling
when the pseudo-scalar is light.  Furthermore, our result for matching
the DAMA allowed region is mostly independent
of the light Higgs mass or the squark masses.  Therefore, 
the rest of the superpartner masses can arrange themselves to produce
large enough loop corrections to satisfy the Higgs mass constraints without 
significantly affecting our final result.

\section{Conclusions}

We complement previous efforts to demonstrate the compatibility of
supersymmetry and the recent DAMA annual modulation signal
\cite{ellis00}-\cite{bottino00jan}
by showing that the wino-like LSP from AMSB
models  are also compatible with the DAMA signal.  Furthermore, the
AMSB models lead to detectable event rates for a large volume of SUSY
parameter space through the tuning of $\tan\beta$, $m_\squark$, and
$m_A$.
The AMSB-inspired models produce nearly
identical event rates if a higgsino is the LSP.  This
is due to the dominance of Higgs exchange and the 
$\tilde H_i \leftrightarrow \tilde W$  symmetry in the
$-\frac{1}{\sqrt{2}} g_2 H_i \tilde{H}_i \tilde{W}^3$ operators.

mSUGRA and AMSB models differ by about an order of magnitude in the event
rate when mSUGRA produces a bino-like LSP, due mainly to the differences of
the SU(2) and U(1) gauge couplings.  The wino-like
neutralino's event rate is always higher than the bino-like
neutralino's for all of parameter space.

Although the WIMP mass and cross-section implied by DAMA's recent
observations are confirmed to be compatible with mSUGRA and AMSB models,
the necessary parameters are near the edge of what is ruled out by
other experiments, such as collider physics.  For wino and higgsino
LSPs, the masses can be rather light (certainly lower than 80 GeV)
without running into collider constraints
because it is so difficult to find winos and higgsinos at 
colliders~\cite{Chen:1996yu}-\cite{Gunion:2000jr}.

\noindent
\emph{Acknowledgements}:  B.M. thanks S. Mrenna for useful discussions.
This work was supported by the Department of Energy and the Alfred P.
Sloan Foundation.

\section*{Appendix:  MSSM Coupling Constants}

For the Higgs-fermion-fermion vertices, using the notation of Drees
and Nojiri (Ref \cite{drees93oct}):
\begin{eqnarray}
c_{h\lsp\lsp} &=& \frac{1}{2} (g_2 N_{12} - g_1 N_{11})
(\sin\alpha N_{13} + \cos\alpha N_{14}) \label{hXX} \\
c_{H\lsp\lsp} &=& \frac{1}{2} (g_2 N_{12} - g_1 N_{11})
(-\cos\alpha N_{13} + \sin\alpha N_{14}) \label{HXX} \\
c_{hdd} &=& \frac{g_2}{2m_W}\frac{\sin\alpha}{\cos\beta} \\ 
c_{Hdd} &=& -\frac{g_2}{2m_W}\frac{\cos\alpha}{\cos\beta} \\
c_{huu} &=& -\frac{g_2}{2m_W}\frac{\cos\alpha}{\sin\beta} \label{huu} \\
c_{Huu} &=& -\frac{g_2}{2m_W}\frac{\sin\alpha}{\sin\beta}
\end{eqnarray}
\noindent
where $\alpha$ is the Higgs mixing angle.

For the $Z\lsp\lsp$ vertex: \beq O''_R = \frac{1}{2} \left[ |N_{14}|^2
- |N_{13}|^2 \right] \eeq

For the $\squark q \lsp$ vertices:
\begin{eqnarray}
a_{\squark_1} &=& \frac{1}{2} [\cos\theta_\squark(X_{q0} + Z_{q0})
+\sin\theta_\squark(Y_{q0} + Z_{q0})] \\ a_{\squark_2} &=&
\frac{1}{2} [-\sin\theta_\squark(X_{q0} + Z_{q0})
+\cos\theta_\squark(Y_{q0} + Z_{q0})] \\ b_{\squark_1} &=&
\frac{1}{2} [\cos\theta_\squark(X_{q0} - Z_{q0})
+\sin\theta_\squark(-Y_{q0} + Z_{q0})] \\ b_{\squark_2} &=&
\frac{1}{2} [-\sin\theta_\squark(X_{q0} - Z_{q0})
+\cos\theta_\squark(-Y_{q0} + Z_{q0})]
\end{eqnarray}
\noindent
where $\theta_\squark$ is the squark mixing angle of left and right
squarks into physical squarks and
\begin{eqnarray}
X_{q0} &=& -\sqrt{2} g_2 [T_{q3} N_{12} - \tan\theta_W (T_{q3} - e_q)
N_{11}] \\ Y_{q0} &=& \sqrt{2} g_2 \tan\theta_We_q N_{11}
\\ Z_{q0} &=& \left\{
\begin{array}{ll}
-\frac{g_2 m_u N_{14}}{\sqrt{2}\sin\beta m_W} & \textrm{for up-type
quarks} \\ -\frac{g_2 m_d N_{13}}{\sqrt{2}\cos\beta m_W} & \textrm{for
down-type quarks.}
\end{array}
\right.
\end{eqnarray}


\end{document}